\documentclass[aps,prd,twocolumn,nofootinbib,superscriptaddress]{revtex4} 

\usepackage{graphicx}
\usepackage{amsmath,amssymb,amsfonts}
\usepackage[colorlinks,citecolor=blue,linkcolor=blue,urlcolor=blue]{hyperref}
\usepackage{xcolor}

\usepackage[english]{babel}
\usepackage[utf8]{inputenc}
\usepackage[T1]{fontenc}

\usepackage{physics} 
\usepackage[percent]{overpic} 

\newcommand{\be}{\begin{equation}}
\newcommand{\ee}{\end{equation}}

\usepackage[autostyle, english = british]{csquotes}
\MakeOuterQuote{"}

\usepackage[colorinlistoftodos,prependcaption,textsize=footnotesize]{todonotes}

\begin{document}

\title{Diffuse emission from black hole remnants}

\author{Sina Kazemian}
\email{skazemi5@uwo.ca }
\affiliation{Dept.\,of Physics \& Astronomy, Western University, N6A\,3K7, London ON, Canada}
\author{Mateo Pascual}
\email{mpascua@uwo.ca}
\affiliation{Dept.\,of Physics \& Astronomy, Western University, N6A\,3K7, London ON,  Canada}
\author{Carlo Rovelli}
\email{crovelli@uwo.ca}
\affiliation{Dept.\,of Philosophy and Rotman Institute, Western University, N6A\,3K7, London ON, Canada}
\affiliation{Aix Marseille University, Universit\'e de Toulon, CNRS, CPT, 13288 Marseille, France}
\affiliation{Perimeter Institute, 31 Caroline Street North, N2L\,2Y5 Waterloo ON,  Canada}
\author{Francesca Vidotto}
\email{fvidotto@uwo.ca}
\affiliation{Dept.\,of Physics \& Astronomy, Western University, N6A\,3K7, London ON, Canada}
\affiliation{Dept.\,of Philosophy and Rotman Institute, Western University, N6A\,3K7, London ON, Canada}

\date{\small\today}

\begin{abstract} 

\noindent  
At the end of its evaporation, a black hole may leave a remnant where a large amount of information is stored. We argue that the existence of an area gap as predicted by Loop Quantum Gravity removes a main objection to this scenario. Remnants should radiate in the low-frequency spectrum. We model this emission and derive properties of the diffuse radiation emitted by a population of such objects. We show that the frequency and energy density of this radiation, which are measurable in principle, suffice to estimate the mass of the parent holes and the remnant density, if the age of the population is known. 

\end{abstract}

\maketitle 

\section{Introduction}

The possibility that black holes tunnel into long-living remnants at the end of their evaporation  \cite{J.H.MacGibbon1987,Giddings1992a,haggard_black_2015, haggard_quantum-gravity_2015,bianchi_white_2018,Rovelli2018f} has recently been receiving renewed attention \cite{Adler2001,Nozari2008,Inomata2021,DiGennaro2021,Green2021,EslamPanah2020,chen_black_2003,Chen2004}. Here we study the emission that we may expect from such remnants.

As a black hole approaches the end of its evaporation, it enters a Planckian regime, since the curvature of its surroundings reaches the Planck scale. A number of recent results from non-perturbative quantum gravity as well as from classical General Relativity have revived the old idea \cite{J.H.MacGibbon1987,Giddings1992a} that the end of the evaporation could leave long living Planck scale remnants. These results are: {\bf\textit{(i)}} \emph{Classical} General Relativity provides a surprisingly natural model for such remnants: white holes with a small horizon but very large interior \cite{haggard_black_2015}. These are exact  solutions of Einstein's equations. 
{\bf\textit{(ii)}} \emph{Classical} General Relativity allows for the existence of spacetimes where such white holes are in the future of the parent black hole, after a quantum tunnelling transition localised in space and time \cite{haggard_black_2015} (the black to white tunnelling transition is itself an old idea \cite{narlikar_high_1974, hajicek_singularity_2001, ambrus_quantum_2005, olmedo_black_2017}). 
{\bf\textit{ (iii)}}~Non-perturbative calculations in Loop Quantum Gravity (LQG) show that the tunnelling transition is permitted, and increasingly probable towards the end of the evaporation \cite{christodoulou_planck-star_2016,  christodoulou_characteristic_2018,DAmbrosio2021,soltani2021}. {\bf\textit{(iv)}} Planck scale white holes can be stabilised against instability by quantum gravity \cite{rovelli_small_2018}. {\bf\textit{ (v)}} A number of objections that made the remnant idea unconvincing a few decades ago \cite{banks_are_1992, giddings_dynamics_1992, banks_black_1993, giddings_constraints_1994, banks_lectures_1995} have now been shown not to apply to this scenario \cite{bianchi_white_2018}. We also give below a further argument removing previous objections to the idea of remnants. 
 
Thanks to these results, long living remnants are again plausible outcomes for the end of the evaporation. Notice that the alternative idea that black holes could magically pop out of existence is a scenario not directly predicted or supported by any quantum gravity theory, and hard to harmonize with the persistence of the large volume inside evaporating black holes \cite{Christodoulou2015, de_lorenzo_improved_2016}.   

If a black hole ends up tunneling into a white hole, its horizon is not an event horizon because its interior is causally connected with future null infinity. 
On the other hand, causality prevents the black hole from being an ergodic system, because energy cannot freely traverse the horizon in both directions.  For a non-ergodic system, the von Neumann entanglement entropy can be higher than the thermodynamic entropy, because all degrees of freedom can contribute to the first, but they do not necessarily contribute to the second \cite{Rovelli2017e,Rovelli2019a}. Hence the von Neumann entanglement entropy across the horizon of a black hole can remain high even when the Bekenstein-Hawking entropy decreases throughout the evaporation. This is a possibility which is manifestly distinct from the popular strong version of the holography "dogma" \cite{Maldacena2020}, but does not contradict any known physics. In this scenario, the interior state of the black hole can still be highly entangled with the emitted Hawking radiation when the black hole reaches Planck mass and this is sufficient to purify the Hawking radiation. Information has no reason to start escaping at Page time and remains stored in the (vast) hole's interior when the horizon tunnels from black to white.  

But eventually information has to come out, before the final dissipation of the white hole. The minuteness of the white hole horizon's area and energy implies that this can only happen slowly \cite{preskill_black_1992}.  Bringing out a large amount of information involving only little energy is what gives rise to the low energy radiation that we model here. 

Here we study the characteristics of this radiation and in particular the diffused background produced by a population of remnants. We use a crude model, with only a few parameters, that provides a quantitative estimate of the property of this radiation as a function of the original mass $m$ of the black hole at its time of formation. We stress the fact that what we give is not a first-principles derivation of the radiation profile (as Hawking did with the black hole radiation), but only an estimate of its properties based on a simple modelization and exploiting basic conservation principles. 

Our results confirm previous estimates that black hole remnants take a long time to radiate ($t \sim m^4$ in natural units). We also estimate the frequency and amplitude of this diffuse background radiation. These determine the number density of the remnants, as a function of the characteristic mass and time of formation of the parent black hole population. We also discuss the quantum field theoretical picture of the remnant-radiation coupling, showing that a key result in non-perturbative Loop Quantum Gravity resolves a potential objection to this scenario. 
 
An interesting possibility is that a population of small long-living black hole remnants produced by primordial black holes \cite{Carr1974a}, born either in the early universe or before a big-bounce, could be a component of dark matter \cite{Rovelli2018f,chen_black_2003,Chen2004,Green2021,Inomata2021,DiGennaro2021,EslamPanah2020}. At first sight, Planck size white holes stabilized by quantum gravity seem  natural cold dark matter candidates, since they interact almost only gravitationaly, behave like a rarefied dust of micro-gram size grains, and their existence is predicted by current fundamental theory (General Relativity plus quantum tunnelling) without any  new physics. A preliminary exploration of the version of this scenario with post big-bang black holes \cite{derome_improved_2003} finds difficulties with it, but not with the bouncing scenario. A white-hole-remnant dark matter component  would be hard to detect directly. Perhaps the background low frequency emission we describe here could help in this regard. Here, however, we only study the physics of the emission by a population of these objects in flat space: we do not delve into any question concerning its cosmological evolution or concrete possibility of observation.

\section{The model}

The hypothesis of the model we study is the following. A black hole of initial mass $m$ evaporates via Hawking evaporation, leaving behind a remnant of Planckian mass which contains an amount of information that is sufficient to purify its Hawking radiation, namely of order 
\be
S\sim \frac{A}4= 4\pi m^2
\label{S}
\ee
 in natural units $\hbar=G=c=k=1$. Here $A$ and $m$ are the area of the horizon and the mass of the hole {\em at its formation}, not to be confused with area and mass at the time of tunnelling, which are presumably approximately Planckian. Our assumption is that the information associated to $S$ is later emitted in the form of radiation. 

Let $\tau_W$ be the lifetime of the remnant, namely the time lapsed from the black-to-white transition of the hole to the moment the white hole has completely dissipated. By causality, at this final time the radiation is spread non-uniformly over a sphere of radius $L=\tau$ (remember that we have $c=1$). The radiation is emitted radially, hence its propagation direction is everywhere in a single direction.  Developing an idea hinted at in \cite{preskill_black_1992}, we model the radiation along a single radius of this sphere as a uniform one-dimensional gas of photons in equilibrium. This is a crude approximation, but presumably sufficient to give us orders of magnitude. 

The total energy $E$ available for this gas of photons is only that of the mass of the remnant, which is of the order of the Planck mass, namely  
\be
E\sim 1
\ee
in natural units. Its total entropy, needed to purify the Hawking radiation, is \eqref{S}, distributed over the length $L$.

A standard derivation, recalled in the appendix for completeness, shows that the entropy $S$ and energy $E$ of a one dimensional photon gas of temperature $T$ in a space of length $L$ are \cite{schwabl_statistical_2006,skobelev_entropy_2013}
\be
S=\frac{2\pi}{3}LT, \ \ \ \  E=\frac16 L T^2.
\ee
Inverting these two relations we find 
\be
L= \frac{3 \text{S}^2}{8 \pi ^2 \text{E}}=6m^4,\ \ \ \ T= \frac{4 \pi 
   \text{E}}{\text{S}}=\frac1{m^2}. 
   \label{tempe}
\ee
From the definition of $L$ we have that the lifetime of the white hole is therefore
\be
\tau_W\sim 6 m^4
\label{tauB}
\ee
which matches previous estimates of the time needed to release the information contained inside the remnant
\cite{preskill_black_1992,Marolf2017,rovelli_small_2018,bianchi_white_2018}.  
The estimate indicates that the temperature of the white hole is much lower than the Hawking temperature of the parent black hole which evolves as $1/m(t)$ ($m(t)$ is the mass of the black hole at time $t$, as opposed with $m$ being the mass of the black hole when it first forms), thus increasing during the evaporation. Notice that the model corrects the infinite explosion predicted by Hawking\footnote{The title of his 1974 Nature paper \cite{HAWKING1974} was \emph{"Black hole explosion?"}}: the black hole temperature grows only up to the Planckian value, then drops abruptly to the low value \eqref{tempe}.

At this point, we can use the crude model we've constructed describing a gas of photons in equilibrium to furthermore estimate the total number of photons emitted by the remnant. At equilibrium, the peak frequency of the (Planckian) distribution of photons is
\be
\nu= \alpha \;T = \frac{\alpha}{m^2}
\label{nu}
\ee
where $\alpha\sim 2.82$. The derivation of the constant $\alpha$ is reported in the appendix for completeness.\footnote{If the frequency is low enough, then free-free absorption makes interstellar gas opaque \cite{King1980}. The Milky Way shows a spectral turnover at 3 MHz attributed to free-free opacity \cite{Lacki2012}. We thank an anonymous referee for pointing this out.}. 

In natural units, the relation between the energy $\epsilon$ of a single photon and its frequency $\nu$ is of course $\epsilon =\nu$, hence we can derive the total number of photons emitted by the remnant of a black hole of initial mass $m$ to be
\be
N_{\gamma}=\frac{E}{\epsilon} = \frac{m^2}{\alpha}.
\label{number of photons emitted}
\ee

Since the emission is radial, the gas of photons at the end of the dissipation process would span a ball of volume $V=\frac43\pi  L^3$. At this time, the average energy density in this volume is therefore
\be
\rho_o=\frac{E}{\frac43\pi L^3}=\frac{1}{288\,\pi } m^{-12}
\ee
and the average photon density 
\be
n_\gamma=\frac{N_\gamma}{\frac43\pi L^3}=\frac{1}{ 288\,\pi \alpha}m^{-10}. 
\ee

Let us now consider a population of remnants uniformly distributed in space with number density $\Omega$, resulting from the evaporation of black holes which all formed at the same time and with the same mass. Neither the energy emitted by a single remnant nor the density of emitted photons are uniformly distributed in space, but the ensemble of all photons emitted by a uniform population of remnants is indeed a uniformly distributed bath of radiation on cosmic scales.

The total energy emitted at the end of the process must be equal to their total initial mass.  Since they have unit mass (in natural units), this is equal to their total number. Hence, the energy density of the radiation $\rho_{tot}$ at the end of the process is equal to the initial number density of remnants,
\be
\rho_{tot}=\Omega
\ee
and the total photon number density (due to the entire population) is
\be \label{total_number_density}
n=\Omega\,N_\gamma=\Omega\,\frac{m^2}{\alpha}\,.
\ee
In a cosmological context, the same result holds, but $\Omega$ and equation \eqref{total_number_density} express comoving number densities instead.

\subsection{\label{sec:level2-1} Linear emission}

Consider a population of black holes formed at a time $t=0$, with mass $m$ and uniformly distributed in space. Assume that they all evaporate around time $\tau_B\sim m^3$ as predicted by Hawking radiation theory, and survive as white hole remnants for a time $\tau_W$ as in \eqref{tauB}. Between times $\tau_B$ and $\tau_B+\tau_W$, they emit a steady radiation as described above. Assuming $m\gg1$, we can approximate $\tau_B+\tau_W\sim \tau_W$. What is the radiation observed by an observer at time $t$?

A fist approximate answer can be given assuming that the radiation is emitted steadily in time. For $t<\tau_B$, there is none.  For $\tau_B<t<\tau_B+\tau_W$, the observer will receive only the radiation emitted by the remnants within a distance $r<(t-\tau_B)$ because radiation emitted by more distant remnants has not had enough time to reach the observer. Radiation emitted at a distance $r$ is diluted by distance by a factor $1/r^2$, but the number of emitters at this distance is proportional to $r^2$, hence the radiation received is proportional to $r<(t-\tau_B)$. (Recall that we are working on flat space. The result can  be extended to a cosmological context, but we do not do so here.) For the same reason, when $t>\tau_B+\tau_W$ and the remnants have completed their emission, the radiation received remains constant in time. That is to say, the radiation density changes in time as
\be
\rho(t)\left\{\begin{array}{lll}
   = 0& {\rm for} & t<m^3, \\ 
    = \big(\frac{t-\tau_B}{\tau_W-\tau_B}\big) \Omega & {\rm for} & m^3<t<6m^4, \\ 
    =\Omega & {\rm for} & t>6m^4. 
\end{array}\right.
\label{array linear emission}
\ee
In other words, the process is a steady (linear in time) transformation of dust into radiation, on a $m^4$ timescale. As we see below, however, quantum mechanics corrects this result. 

\begin{figure}[t]
\includegraphics[width=0.6\columnwidth]{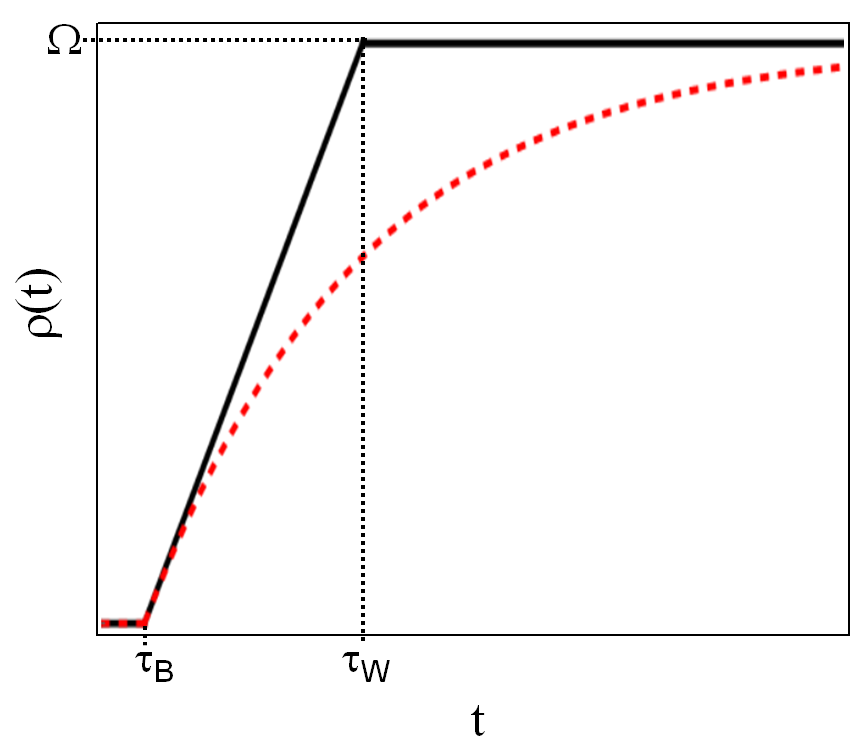}
\caption{Background white hole radiation as a function of time. The solid black line represents a classical linear emission while the dashed red line represents a quantum emission.}
\label{fig:energy density}
\end{figure}

\subsection{\label{sec:level2-2} Quantum description of the emission}

The energy of a white hole is related to the area of its horizon. A continuous, steady emission of energy from each white hole, as the one described above, would imply a continuous decrease of the white hole horizon area, even below the Planck area. According to LQG, however, any physical area is quantized, with the minimum non-zero eigenvalue (the "area gap") being of the order of the Planck area $A_{Pl}$ \cite{Rovelli1993c,Smolin1994,Rovelli1994a}. Therefore a remnant with near-Planckian mass and area cannot emit an amount of energy smaller than an energy of the order of the Planck energy. That is, it can only make a single quantum leap into radiation. This is analogous to conventional nuclear radioactivity, where the steady emission of a macroscopic sample of radioactive material is realised by individual quantized emissions of its constituent atoms, governed by a probability distribution \cite{martin_nuclear_2019}. 

More precisely, an area gap of the order of the Planck area implies that the energy of the lowest non-vanishing energy states of the remnants is Planckian. Therefore, in first-order perturbative formulation, the only allowed transition with the emission of radiation (which necessarily has energy) is emission of the entire Planck energy of the remnant.   

Let us see, in the language of quantum field theory, what the corresponding vertex describing the transition could be. The essential point is that (black and) white holes have many internal degrees of freedom that reflect their internal structure. A white hole parented by an old black hole evaporated from a initial mass $m$ has an interior capable of holding information compatible with \eqref{S} even if the area of its horizon is small.  A vertex coupling such remnant to a single or a few photons is therefore forbidden by conservation of information (unitarity), because a few photons do not have enough degrees of freedom to match the large number of quantum numbers describing the white hole interior. Just a few photons  cannot carry the entire information that can be stored in the remnant. Hence the only possible transition is a transition $remnant \to \gamma_1...\gamma_n $ to {\em a large number} of low energy photons: \\[1em]
\centerline{~~~~~~~~~~~~~~~~~~\includegraphics[width=0.6\columnwidth]{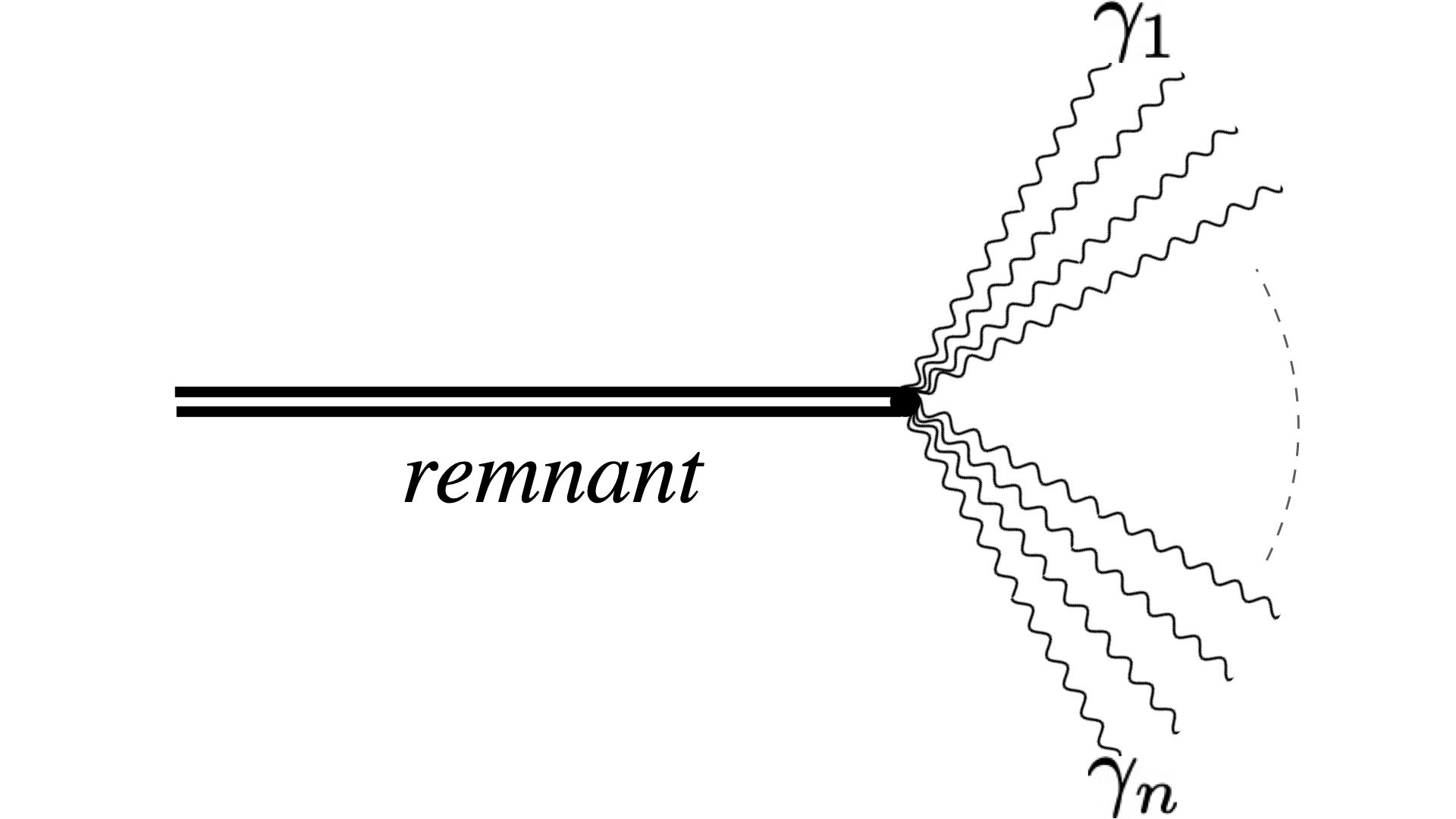}}
\\[1em]

This conclusion is interesting in view of an old objection to the remnant scenario, because of which this scenario was abandoned in the Nineties \cite{preskill_black_1992}. The objection was that the large number of remnant internal states would entail that their production in particle physics experiments would be too easy, hence expected be observed already. Here we see clearly why that conclusion was too quick. The effective vertex responsible for a remnant production would actually have to be 
\\[1em]
\centerline{\includegraphics[width=0.6\columnwidth]{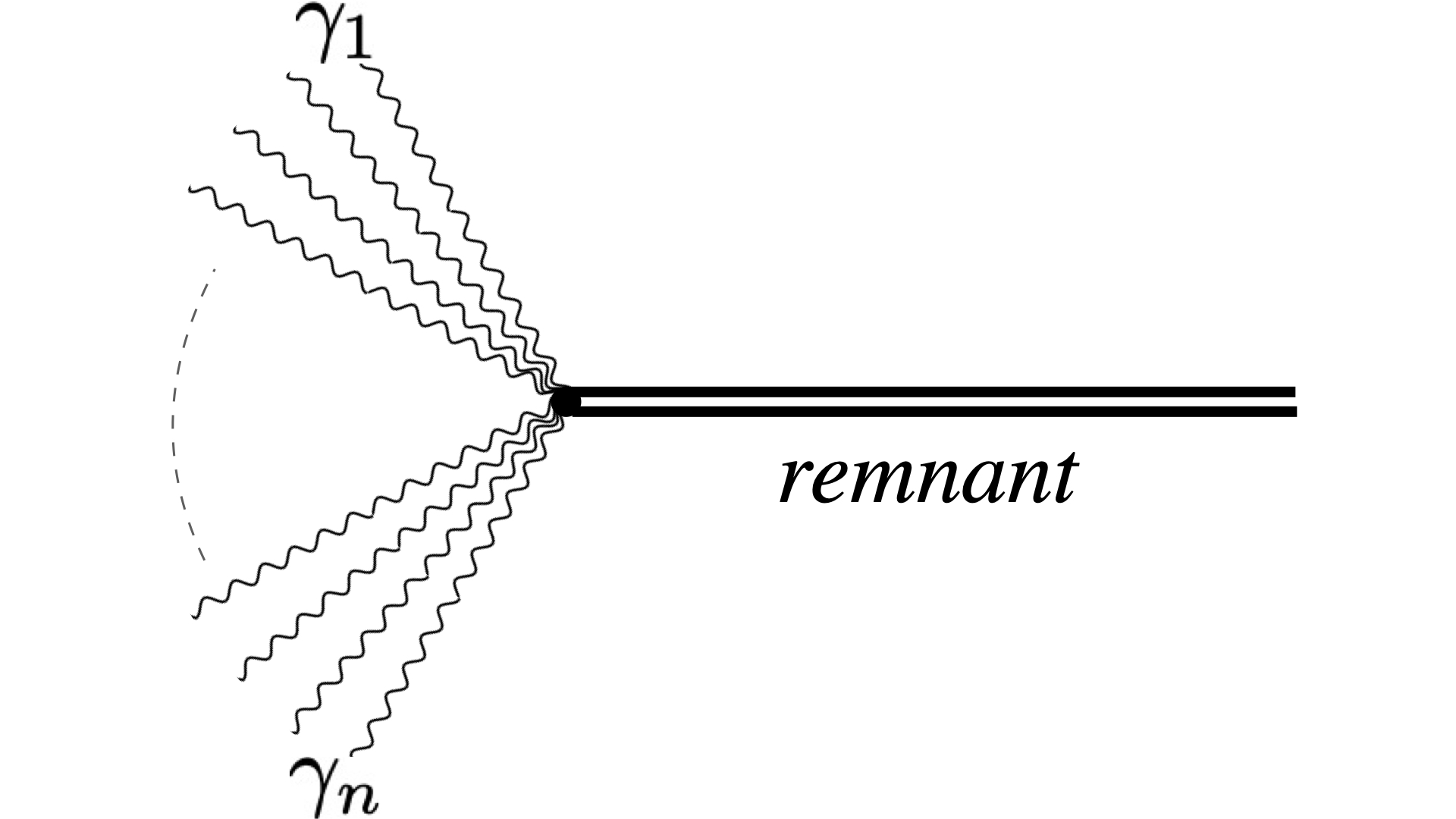}}
\\[1em]
\noindent in order to create a long-living, Planck-size remnant. The number of photons emitted by a single remnant is given in \eqref{number of photons emitted}. If the number of photons coming together is small, they can be highly energetic, but the remnant produced would correspond to a remnant whose parent is a black hole of Planckian size, which is short-lived. The process would not be distinguishable by the standard possibility of collapse predicted by conventional quantum gravity. To produce an actual long living remnant, on the other hand, we need $m$ to be large, and hence we would need to focus {\em a large number of low energy photons}. 

For instance to produce a remnant similar to the one left over from a primordial black hole formed at reheating (see below) the number of low energy photons to focus would be staggering: 
\be
5\cdot 10^{38}<N_{\gamma}<5\cdot 10^{48}. 
\ee
Creating such a remnant in the lab is clearly unlikely due to the huge number of photons required for the process to happen. Therefore not being able to create remnants by the present experimental settings is not a reason to reject the theory of black holes turning into Planck size white hole remnants at the end of their life time cycle.\\ 

Let us see how this quantum effect corrects the expected emission of a population of remnants. If there is a single decay into multiple photons for each remnant, and 
the probability of transition is constant in time, then a \emph{population} of remnants behaves like a radioactive material: the number of remnants, and therefore the total energy density $\rho_{rem}$ of the population of remnants decays  exponentially, starting at $t=\tau_B$, as 
\be
\rho_{rem}(t)=\Omega\ e^{-\lambda(t-\tau_{B})}.
\label{energy density of the quantum emission}
\ee
If the lifetime of the white hole is of order $\tau_{W}$ (by Bohr's correspondence principle), we expect the decay constant to be 
\be
\lambda\sim(\tau_{W}-\tau_{B})^{-1}.
\label{decay constant}
\ee
The energy density of the radiation as a function of time is therefore
\be
\rho_{rad}(t)\left\{\begin{array}{lll}
   = 0& {\rm for} & t<m^3, \\ 
    = \Omega \bigg(1-e^{-\frac{t-\tau_{B}}{\tau_{W}-\tau_{B}}}\bigg) & {\rm for} & t>m^3. 
\end{array}\right.
\label{array quantum emission}
\ee
In Figure \ref{fig:energy density} we have plotted the energy density of the linear emission \eqref{array linear emission} as a black solid line, and the quantum emission \eqref{array quantum emission} as a dashed red line. The two converge in the two limits $\tau_{B}<t\ll\tau_{W}$ and $t\gg\tau_{W}$. 

The energy density $\rho_{rad}$ of the radiation emitted by a population of remnants with current energy density $\rho_{rem}$, generated by parent black holes that all formed a time $t$ in the past with mass $m$, is then obtained by combining equations \eqref{energy density of the quantum emission} and \eqref{array quantum emission}, which gives
\be
\rho_{rad}(t,m) = \rho_{rem} \left[ \exp\! \left( \frac{1-t m^{-3}}{1-6m} \right) -1 \right]. 
\ee

Using \eqref{nu} we can write the mass in terms of the peak frequency of the radiation, and give the energy density in radiation as
\be
\rho_{rad}(t,\nu) = \rho_{rem} \left[ \exp\! \left(\frac{1- t\, (\nu/\alpha)^{3/2}}{1-6\sqrt{\alpha/\nu}}\right) -1 \right] . 
\ee

\section{Dimensionful estimates for primordial holes}

In the cosmological standard model, primordial black holes may have formed at reheating. To get a sense of the characteristic nature of the diffuse radiation remnants may emit, we estimate its parameter in this  simplest case.  We use here a rough model, which neglects the effect of cosmological expansion. 

For such a population of primordial black holes, we can approximate $t$ in the above formulas to be the Hubble time $t_H$. Notice that this approximation allows us to deduce the density of an otherwise dark population of remnants just from the observation of the emitted radiation. 
(In other cosmological scenarios,
in particular in bouncing models 
\cite{ashtekar_loop_2015,brandenberger_bouncing_2017,ijjas_bouncing_2018},
$t$ can be larger.)

Restoring physical units, and denoting the Plank mass, energy, frequency and time as $m_{Pl}$, $E_{Pl}$, $\nu_{Pl}$, $t_{Pl}$ we find that a population of primordial black holes of formation mass $m$ and comoving number density $\Omega$  gives rise to remnants emitting diffuse radiation with density 
\be
\rho_{rad}(m) = \Omega E_{Pl}\bigg(1-e^{\frac{1-t_H/t_{Pl}\,(m/m_{Pl})^{-3}}{1-6m/m_{Pl}}}\bigg)
\ee
and frequency given by \eqref{nu}, namely 
\be
\nu=\left(\frac{m}{m_{Pl}}\right)^{\!\!-2}\nu_{Pl}.
\ee

If we are in the era where this radiation forms, we must have $\tau_B<t_H<\tau_W$, which is to say 
\be
(m/m_{Pl})^3<t_H/t_{Pl}<6(m/m_{Pl})^4.
\ee
Since $t_H \sim 10^{61}t_{Pl}$ this gives the approximate mass range
\be
10^{15}m_{Pl}<m<10^{20}m_{Pl}.
\ee
The model is thus  entirely determined by a single parameter or order of unity, that can be taken to be 
\be 
x=\log_{10}(m/m_{Pl})\in(15,20].
\ee 
And the relevant quantities are 
\begin{eqnarray}
m &=& 10^{x-5}{\rm g},  \\
\nu &=& 5\cdot 10^{-2x+42}{\rm Hz}\\
\rho_{rad}(x) &=& \rho_{rem} \left[ \exp\! \left(\frac{10^{61}-10^{3x}}{10^{4x}-10^{3x}}\right) -1 \right] 
\end{eqnarray}
This is a mass range
\be
10^{10}{\rm g}<m<10^{15}{\rm g},
\ee
and a frequency range 
\be
5\cdot10^{12}{\rm Hz}>\nu>5\cdot10^{2}{\rm Hz}. 
\ee
The ratio of the radiation density to the total density of remnants and radiation, as a function of $x$, is shown in Figure 2.
Notice that remnants originating from parent black holes in the mass range of $10^{10} g <m<2\cdot 10^{10} g$ have emitted most of their energy in radiation as of today, while remnants originating from more massive black holes, $4\cdot 10^{10}<m<10^{15}$ have emitted close to zero. This is because of the long lifetime of white hole remnants which is in the order of $\tau_{W}=6m^4$.      
\begin{figure}[t]
\includegraphics[width=.6\columnwidth]{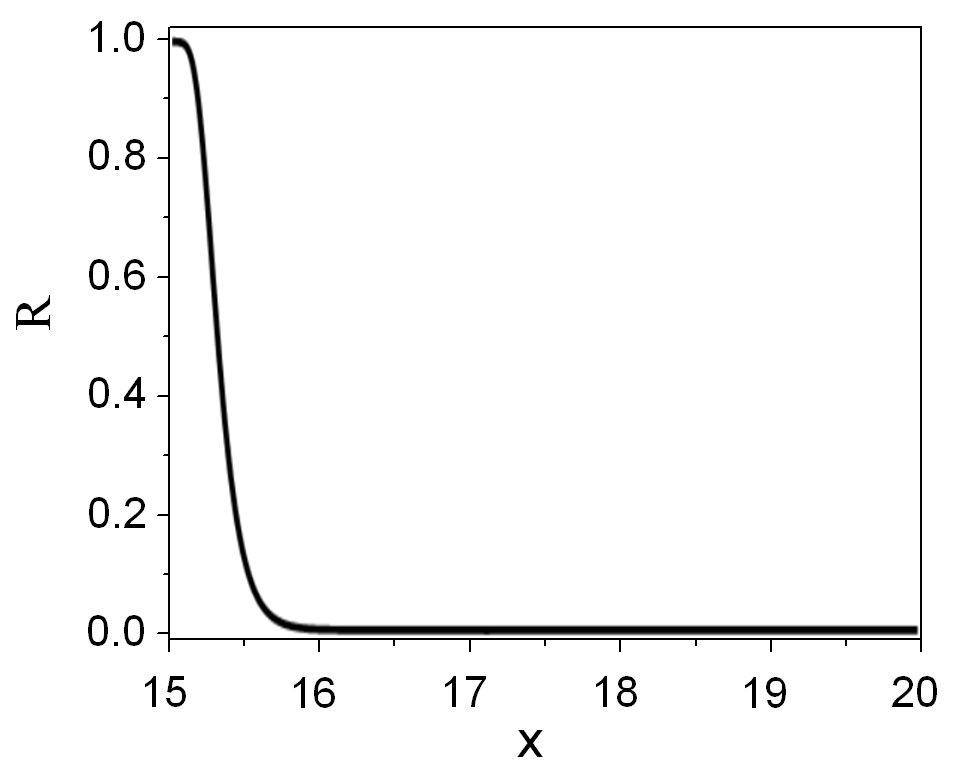}
\caption{Ratio of radiation to total mass as function of the single parameter of the model $x$.}
\end{figure}

\section{Conclusions}

We have shown that a quantum gravity effect, namely the quantization of the area, removes a main objection to the possibility that the end of a black hole's evaporation leaves behind a Plank scale remnant. The area gap gives a wide forbidden energy region, because of which the remnant can only couple to a large number of low energy photons. This makes remnant production hard to achieve in a lab. 

We have estimated that a diffuse radiation bath peaked at frequency $\nu$ with energy density $\rho_{rad}$ can account for a population of white hole remnants descending from black holes formed a time $t$ in the past, with mass 
\be
m=10^x m_{Pl}
\ee
where 
\be
x=-\frac12\ \log_{10}\frac{\nu}{ \nu_{Pl}},
\ee
and density
\be
\rho_{rem} = \rho_{rad} \left[ \exp\! \left(\frac{1- t\, (\nu/\alpha)^{3/2}}{1-6\sqrt{\alpha/\nu}}\right) -1 \right]^{-1} .
\ee

For black holes formed a Hubble time in the past, this approximately becomes
\be
\rho_{rad}(t,\nu) = \rho_{rem} \left[ \exp\! \left( \frac{1 - 10^{61 - 3x}}{1 - 6\cdot 10^{-x}} \right) -1 \right] ,
\ee
where $x$ can be measured directly from the frequency of the diffused radiation. These results are in flat spacetime, and can be easily generalised to an expanding cosmological setting. 

To evaluate the hypothesis that remnants might form a component of dark matter, these results must be corrected by taking into account the cosmological evolution. For black holes formed after the big bang in the primordial universe, there are strong constraints \cite{barrau_closer_2021}. In other cosmological scenarios such as big bounce or matter bounce scenarios white hole remnants might account for a significant portion of dark matter.

\vskip2mm
\centerline{$***$}
\vskip2mm

\noindent
{\bf Acknowledgments}
We thank Emily Adlam, Patrick Fraser, Pietropaolo Frisoni, Yichen Luo and Farshid Soltani for participating in many discussions about this work. 
This work was supported by the QISS JFT grant 61466.
Research in FV's group at Western University is supported by the Canada Research Chairs Program and by the Natural Science and Engineering Council of Canada (NSERC) through the Discovery Grant "Loop Quantum Gravity: from Computation to Phenomenology".  
We acknowledge the Anishinaabek, Haudenosaunee, L\=unaap\'eewak and Attawandaron peoples, on whose traditional lands Western University is located.%

%
\appendix

\section\! 

Here we derive of the entropy and the energy of a one-dimensional photon gas emitted by a remnant. We also derive the frequency of the photon gas as a function of temperature. 

The Entropy of the photon gas can be written as \cite{skobelev_entropy_2013}:
\begin{equation}
    S= \frac{k_B^2}{\hbar}\delta^n\frac{4(n+1)(n-1)!\zeta(n+1)}{\Gamma(\frac{n}{2})},
\label{entropy of photon gas}
\end{equation}
where we have:
\begin{equation}
    \delta=\frac{LT}{2\sqrt\pi},
\label{constant alpha def}
\end{equation}
where T is the temperature and L is the distance traveled by the photon gas. The Gamma function is equal to:
\begin{equation}
    \Gamma(n)=(n-1)!,
\label{Gamma def}
\end{equation}
and the Riemann series function is equal to:
\begin{equation}
    \zeta(n)=\sum_{k=1}^{\inf}\frac{1}{k^n}.
\label{zeta def}
\end{equation}
We assume a one-dimensional photon gas radiating from the remnant and put $n = 1$. The entropy of a one-dimensional photon gas is equal to:
\begin{equation}
    S = \delta\frac{8\zeta(2)}{\Gamma(\frac{1}{2})} = \frac{2\pi}{3}LT.
\label{entropy of photon gas 2}
\end{equation}
The energy of a radiating photon gas can be written as:
\begin{equation}
    E=T t^{n}\frac{4n!\zeta(n+1)}{\Gamma(\frac{n}{2})},
\label{Energy}
\end{equation}
where for n = 1 the energy is equal to:
\begin{equation}
    E=\frac{1}{6}LT^{2}.
\label{Energy 2}
\end{equation}

To calculate the frequency of the photons emitted by the remnant we assume the system under study to behave like a black body radiation and write a Planck distribution for it. The energy density is then equal to \cite{schwabl_statistical_2006}:
\begin{equation}
    u(\nu ,T)=\frac{2h\nu^{3}}{c^2}\frac{1}{e^{h\nu/ k_B T}-1}.
\label{The energy density with a Planck distribution}
\end{equation}
By taking the derivative of the energy density with respect to frequency and putting it equal to zero we find the peak frequency as following:
\begin{equation}
    \frac{h\nu}{k_B T}\frac{e^{h\nu/ k_B T}}{e^{h\nu/ k_B T}-1}-3=0.
\label{The energy density derivatuve with respect to the frequency}
\end{equation}
By changing the variable as $h\nu/k_B T=\alpha$ we have:
\begin{equation}
    (\alpha-3)e^{\alpha}+3=0,
\label{The energy density derivatuve written with variable alpha}
\end{equation}
where we get:
\begin{equation}
    \alpha\approx 2.82.
\label{Derivation of alpha}
\end{equation}
The peak frequency is therefore equal to:
\begin{equation}
    \nu_{peak}=\frac{k_B}{h} \alpha T.
\label{The peak frequency}
\end{equation}
where the maximum number of emissions happen at this frequency.

\bibliographystyle{utcaps}
\bibliography{cosmo,library2}

\end{document}